\documentclass[pra,letterpaper,onecolumn,showpacs,superscriptaddress]{revtex4} 
\usepackage{graphicx,amsmath,amssymb,amsfonts,latexsym,color,dcolumn,bm} 
\begin{document} 
\title{Development of a high sensitivity torsional balance \\
for the study of the Casimir force in the 1-10 micrometer range}

\author{Astrid Lambrecht} 
\affiliation{Laboratoire Kastler-Brossel,Universit\'e
Pierre et Marie Curie,Campus Jussieu,F-75252
Paris, France}

\author{Valery V. Nesvizhevsky} 
\affiliation{Institute Laue-Langevin, 6 rue Jules Horowitz, F-38042
 Grenoble, France}

\author{Roberto Onofrio} 
\affiliation{Department of Physics and Astronomy,Dartmouth
 College,6127 Wilder Laboratory,Hanover,NH 03755,USA}

\affiliation{Dipartimento di Fisica
``G. Galilei'',Universit\`a di Padova,Via Marzolo 8,Padova 35131,Italy}

\author{Serge Reynaud} 
\affiliation{Laboratoire Kastler-Brossel,Universit\'e
Pierre et Marie Curie,Campus Jussieu,F-75252
Paris, France}

\date{\today}

\begin{abstract} 
We discuss a proposal to measure the Casimir force in the parallel
plate configuration in the $1-10\mu$m range via a high-sensitivity torsional balance.
This will allow to measure the thermal contribution to the Casimir force
therefore discriminating between the various approaches discussed so far.
The accurate control of the Casimir force in this range of distances
is also required to improve the limits to the existence of
non-Newtonian forces in the micrometer range predicted by unification
models of fundamental interactions.
\end{abstract}
\pacs{04.80.Cc, 06.60.Sx,11.10.Wx, 12.20.Fv} 
\maketitle

\section{Introduction}

The search for deviations from Newton's gravitation law has been a
recurrent issue for the last three decades. 
Initially motivated by the possibility for deviations from standard 
gravity due to new forces with couplings of order of the gravitational
one \cite{ref8}, this search has more recently been encouraged by
unification models which predict the existence of forces which can be
up to $10^5$ times stronger than gravity with ranges between $1\mu$m
and $100\mu$m \cite{ref9,Adelberger03}. 
Even if its results have not met the initial hopes of observing
a ``fifth force'', this search has generated an impressive progress of 
tests of the gravitation law in the laboratory or in the
solar system. It has led to a largely improved knowledge of this law 
narrowing the remaining open windows for new hypothetical forces.

The hypothetical extra-gravitational force is often parametrized
by a Yukawa range $\lambda$ and a coupling strength $\alpha$
such that the corresponding potential is: 
\begin{equation} 
V_{\rm Newton}(d) + V_{\rm Yukawa}(d)= 
-\frac{G M_a M_b}{d} \left( 1 + \alpha e^{-d/\lambda}\right) 
\end{equation}
The Newton and Yukawa potentials have been written for two point masses
$M_a$ and $M_b$ at a distance $d$ from each other, and the coupling 
strength is defined with respect to Newtonian gravity.
The current limits in the $\left(\lambda,\alpha\right)$
plane (see for instance \cite{Coy03}), summarize the considerable 
progress achieved during the last decades, thanks to a variety of 
laboratory experiments and solar system observations. At the same 
time, windows remain open for deviations of standard
gravity in the submillimeter range or for scales larger than the size 
of planetary orbits \cite{Jaekel05ijmp}.
In this paper, we focus our attention on the submillimeter window.

The accuracy of short range tests has recently been much improved 
for Cavendish experiments performed at smaller distances.
The best limits for distances of the order of $\simeq 100\mu$m have been recently
obtained by the group of Adelberger at the University of Washington using
torsion balances and rotors \cite{ref12,ref13,ref13b}. Similar experiments 
have been performed with microresonators in the groups
at Boulder \cite{ref14} and Stanford \cite{ref15} and they have started
to explore distances below $100\mu$m, where however difficulties arise from the stringent requirements to maintain the surfaces  parallel during the rotation. 
For even lower distances, of the order or smaller than one micrometer, the hypothetical 
new forces have to be measured against a large background coming from the Casimir force \cite{ref19}. 
The latter has been measured with increasing accuracy
during the last years by various groups using atomic force microscopes or
microresonators whose motion is monitored by means of
capacitively or optically coupled displacement transducers 
\cite{ref28,ref73,73b,73c,ref74,ref49,ref50,ref56,ref36,ref75,ref75b}.

At this point, it is worth emphasizing that the theoretical predictions are the most promising in the $10\mu$m range \cite{ref16,ref17,ref18}. It is therefore important
to design new experiments aiming at the detection of forces acting in this distances range, where the Casimir force is still the main known background. 
It is the purpose of this paper to describe a proposal for a high precision
study of the Casimir force at distances from a few $\mu$m to $10\mu$m. 
The challenge is to be able to measure the Casimir force which becomes weaker 
and weaker when the distance is increased. On the other hand some of the main corrections
known to endanger the accurate measurement of the Casimir force 
are expected to be more easily controlled at large distances.
Meeting this challenge would allow to bring new information on 
problems still present in the theory of Casimir force 
\cite{ref20,ref21,ref22,ref23,ref24,ref24b,ref25},
and more generally on the understanding of the properties of quantum vacuum
\cite{ref26,ref24c,ref27}.

Another important motivation is coming from open questions in cavity quantum 
electrodynamics, in particular the issue of the finite temperature corrections to 
the Casimir force. This correction is for room temperature particularly significant at large distances, since only then the cavity formed by the two mirrors can sustain electromagnetic
field modes filled with a non-negligible number of
thermal photons \cite{Lifshitz56,Mehra67,Brown69,Schwinger78}. 
The only experiment which has investigated the Casimir force at distances
up to about $8\mu$m in a plane-spherical configuration \cite{ref28},
has shown no evidence of the thermal contribution
which was yet expected to be visible at such distances.
The discrepancy can be interpreted in various manners, going from an 
overestimation of the accuracy claimed in the first letter 
\cite{ref29,ref30,ref31,ref31b} to the much debated possibility 
that previous calculations have mistreated the thermal contribution
for dissipative mirrors \cite{ref32,ref33,ref34,ref60,ref67,ref63,ref68,ref66,ref35}.
Therefore the interplay between
thermal and quantum fluctuations needs to be studied more closely. Obtaining an unambiguous experimental output for the value of the Casimir force at large distances could decide between the contradictory models used by different authors.

Another experiment, performed at the University of Padova, has explored 
the Casimir force at room temperature in a parallel plate configuration \cite{ref36} with
distances up to $3\mu$m. In this experiment, the experimental
accuracy of 15$\%$ was not sufficient to measure the amplitude of the
thermal corrections, weaker at $3\mu$m than at $8\mu$m. Let us emphasize 
that it is important to test the Casimir force in a parallel plane 
configuration for the following reasons \cite{ref37}. First, the parallel 
plane configuration, originally considered by Casimir, is the only one for 
which an exact analytical expression of the force exists. 
For other geometries considered so far, predictions always rely on
the proximity force approximation based on analogies with the treatment 
of additive forces \cite{ref38,ref39}. Second, for an apparatus of a given
size the parallel plane configuration leads to the largest possible signal, because the entire surface fully contributes to the Casimir force. 
This is to be contrasted with the obvious loss of signal in the
plane-sphere geometry due to the smaller size of the
closest approach region between the two surfaces. 
This argument favors the parallel plane configuration for the purpose of testing 
thermal contribution at large distances, since one important drawback of this 
distance range is the rapid decrease of the amplitude of the Casimir force.
Finally, the purely gravitational force does not depend on distance, in a
parallel plane configuration, apart from modelizable border effects.
It is thus easier to disentangle standard gravity from other possible forces.

The main drawback of the parallel plane
configuration is the need for a careful parallelization of the two surfaces. 
In the experiment carried out at the University of Padova \cite{ref36}, 
this was the major factor limiting the accuracy to a level of 15$\%$. 
The solution proposed in the present paper is to take advantage of the 
recent progress in parallelization techniques obtained in a neutron
experiment performed at the Institute Laue-Langevin (ILL) in Grenoble 
\cite{ref51,ref52,ref53,ref54,ref100}.
This technique, described in more detail below, should allow us to overcome
the limitation associated with parallelization of the plates. 

\section{Estimates of signals and backgrounds} 

In this section, we briefly sketch the estimates of signals and 
backgrounds expected in our configuration.

When evaluated for a null temperature, the Casimir force depends only on the
separation between the two mirrors $d$, their surface $S$, and the two fundamental
constants of relativistic quantum mechanics, the Planck constant
$\hbar$ and the speed of light $c$, as: 
\begin{equation} 
F_{\rm Cas}= \frac{\pi^2 \hbar c}{240} \frac{S}{d^4} 
\end{equation} 
Throughout the paper, we consider the case of a rectangular mirror of size
10cm$\times$12cm. We then find a Casimir force $\simeq 25$nN 
at $L=5\mu$m or $\simeq 1.5$nN at $L=10\mu$m.
In fact, at this distance the force is significantly enhanced by the
presence of room-temperature black-body photons. Assuming the mirrors to be 
perfectly reflecting at all frequencies and both polarizations, we get a 
simple expression which gives a correct value at distances larger than 
5$\mu$m:
\begin{equation} 
F_{\rm thermal}= \frac{\zeta(3) k_B T}{4 \pi} \frac{S}{d^3} 
\end{equation} 
At $T=300$K, this force reaches a value $\simeq 38$nN at $L=5\mu$m or 
$\simeq 5$nN at $L=10\mu$m. More accurate expressions of the thermal 
Casimir force can be found in \cite{ref61,ref62} which also
treat the case of mirrors described by a lossless plasma model.

The real mirrors used in the experiments are actually better described by a Drude 
model taking into account dissipation of the optical response of electrons.
For this situation, the various models used for calculating the thermal Casimir
force predict values lying between the preceding expression of 
$F_{\mathrm thermal}$ and a value smaller by about $50\%$. 
This comes from the fact that some of these models predict the two polarizations, transverse electric (TE) and transverse magnetic (TM), 
to contribute equally to the force whereas others predict the contribution of the TE polarization
to vanish for dissipative mirrors
\cite{ref32,ref33,ref34,ref60,ref67,ref63,ref68,ref66,ref35}.
Therefore, at distances between 5 and 10$\mu$m, even an accuracy of a
few percent would be enough to bring new interesting information about the
temperature corrections to the Casimir force. 
This point is discussed in more details in \cite{ref70} (see in particular its 
figure 2) where is shown that the discrepancy between the two families of 
models is particularly emphasized in the plane-plane geometry.

When considering the role of possible backgrounds, it first appears that
spurious electrostatic charges could be a nuisance. For instance, 
a constant stray electric potential of 0.1V would give rise to a force of 
$\simeq 25\mu$N at $L=5\mu$m or $\simeq 5\mu$N at $L=10\mu$m. 
This value is larger by roughly three orders of magnitude than the force we wish to measure. 
This challenge has been faced in all previous experiments through the 
application of counterbias voltages canceling as perfectly
as possible the stray voltage. We have to acknowledge that the solution of 
this problem is more demanding at larger gaps, requiring a control of the stray voltage at a
level better than one part in $10^3$.
Local fields must also be controlled with the greatest care, but this
point is, in contrast to the previous one, less demanding at larger gaps if the 
typical correlation length of patch potentials remains constant \cite{ref55}. 
These points will be taken care of at ILL in Grenoble where the surfaces will be
precisely characterized thanks to locally available surface analysis facilities. 
The spurious electrostatic effects will then be corrected either by applying 
counterbias voltages or via off-line data analysis. The optimal 
distance for the measurement will be determined by a kind of balance between the 
control of bias and locally varying spurious fields. 
Concerning border effects related to the finite surface area of the
plates, we can estimate their influence using recent 
numerical results \cite{refholger} according to which the change 
in the Casimir energy can be taken into account by introducing an effective 
surface area. In the case of a scalar field, the effective area is 
related to the geometrical area $S$ and its perimeter $C$ 
by the relation $S_{\rm eff} \simeq S+0.36 d C$. 
This leads to a relative correction $\Delta F_{\rm Cas}/F_{\rm Cas}
\simeq 0.12 C d/S$, directly proportional to the plate separation $d$. 
For square mirrors with a surface area of the order $10 \times 10$ cm$^2$ at a 
distance of $1 \mu$m we get $\Delta F_{\rm Cas}/F_{\rm Cas} \simeq 4.8
\times 10^{-6}$. For the electromagnetic field the correction should be 
similar, with a the prefactor $0.36$ in the scalar 
case replaced by unity \cite{refholger1}. In any event, even at the largest separation, 
corrections due to the border effects are too small to be detectable. 
 
Finally, it is important to evaluate the contribution expected from
Newtonian gravity and hypothetical Yukawa forces: 
\begin{equation} 
F_{\rm Newton}(d) + F_{\rm Yukawa}(d)=
\frac{G M_a M_b}{d^2} \left( 1 +\alpha \left(1+\frac{d}{\lambda}\right) 
\exp{(-d/\lambda)} \right) 
\end{equation} 
When integrated over two plane parallel plates with densities $\rho$ and
thicknesses $\tau$, this leads to \cite{Long00}: 
\begin{eqnarray} 
F_{\rm Newton}(d) &=& 2 \pi G \rho_a \rho_b S \tau_a \tau_b \nonumber \\ 
F_{\rm Yukawa}(d) &=& 2 \pi G \rho_a \rho_b S \lambda^2
\alpha \exp{(-d/\lambda)} \left[ 1- \exp{(-\tau_a/\lambda)} \right)
\left( 1- \exp{(-\tau_b/\lambda)} \right] 
\label{YukawaF} 
\end{eqnarray} 
For glass plates with a density $\rho_a=\rho_b=3 \times 10^3$ kg/m$^3$ and a thickness
$\tau_a=\tau_b=15 $mm, we find a Newtonian force $\simeq 10$nN which, as already 
stated, does not depend on distance. If this force is measured with
a torsion balance, it could be subtracted by compensating its action
on both arms of the balance, by comparing measurements performed at
different distances, or by using dynamical detection techniques only
sensitive to the spatial gradient of the force.

\begin{figure}[t] 
\begin{center} \includegraphics[width=0.4\columnwidth,clip]{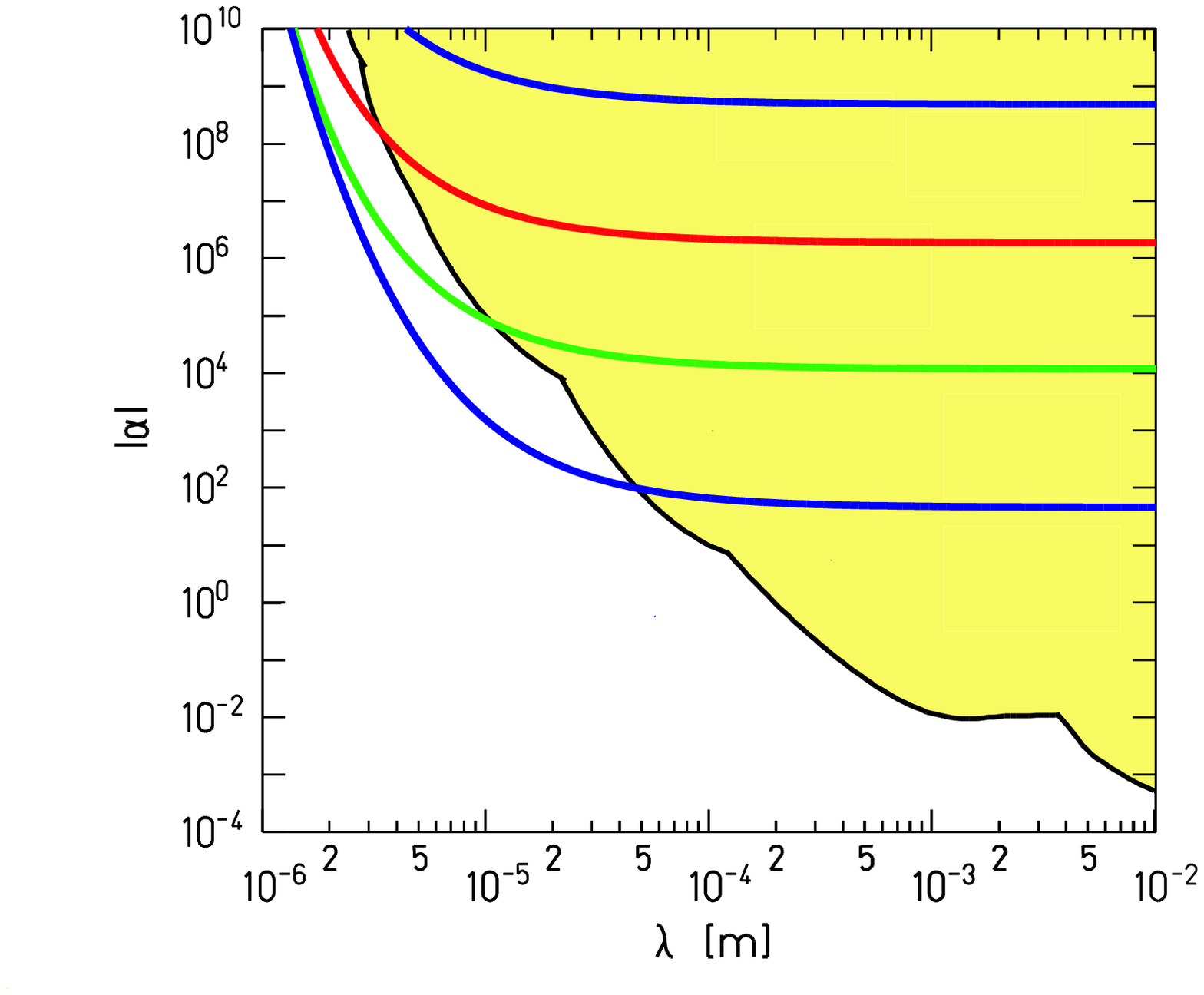} 
\caption{Constraints on Yukawa parameters $\alpha$ versus $\lambda$
deduced from the analysis of the expected sensitivity of the experiment 
proposed in the present paper. 
It is assumed that the resolution reaches the level of 1pN for a force 
measurement at 5$\mu$m, with all systematic effects mastered at the same level.
The 4 curves correspond to different values of the thickness of the gold layer,
respectively $\tau=0.3,1,3,10\mu$m from the top curve to the bottom
one. The shaded region represents the exclusion domain for new hypthetical forces deduced from previous measurements (see Figure 5 in \cite{Adelberger03}).} 
\end{center} 
\label{fignewlim} 
\end{figure}
The Yukawa force (\ref{YukawaF}) is then determined by the parameters of 
the metal layers deposited on the glass plates and by the unknown 
parameters $\lambda$ and $\alpha$.
To give an estimate, we consider the case of gold layers with 
$\rho_a=\rho_b=19.3 \times 10^3$ kg/m$^3$. 
The thickness $\tau_a =\tau_b =\tau$ should be chosen with care since
it plays a key role in the determination of the expression (\ref{YukawaF}): 
the thicker the layer, the larger the hypothetical Yukawa force. 
Note that the estimate of the Yukawa force might be enlarged by adding
the contribution of the glass plate to that of the gold layer. 
However, the metal layer would have to be changed in order to investigate 
the material dependence of the hypothetical Yukawa signal, so that we
prefer to give conservative numbers corresponding to the metal layer alone.

If we consider that we can reach a resolution of the order of 1pN for
the force measurement, with all systematic effects mastered at the same level,
we can translate expression (\ref{YukawaF}) into the following limit on 
the relative Yukawa amplitude $\alpha$ versus $\lambda$: 
\begin{equation} 
\alpha \simeq 532\times \exp \left( \frac{L}{\lambda} \right)
\frac{1}{\lambda^2\left[ 1-\exp \left( -\frac{\tau}{\lambda}\right)\right]^2} 
\end{equation}
The corresponding limit in the $\left(\lambda,\alpha\right)$ plane is shown in Fig. 1 for four different values for the layer thickness, with $\lambda$ spanning from $10^{-6}$m to $10^{-2}$m ($\lambda$ measured in m in the log-log plot).
We compare our limits to the bounds already known at short distances, as 
represented by the black solid line (corresponding to figure 5 in \cite{Adelberger03}).
To fix ideas, the bound obtained for $\lambda=10\mu$m, $L=5\mu$m and
$\tau=10\mu$m is $\alpha\sim 1000$ which would improve the current knowledge by a factor of 100.
Larger values for $\tau$ lead still to better bounds, allowing one either to rule
out theoretical models discussed in \cite{ref16} or to bring new information supporting one of them. Note that smaller values of $\tau$, say a few hundreths of nm, can be used in the 
first stage of the experiment aiming at the test of the thermal Casimir force.

\section{Torsional balance: design and sensitivity}

For the measurement of the force with large surfaces, we intend to use a torsional 
balance similarly to those used in experiments exploring the equivalence principle 
with the E\"{o}tvos technique (see Refs. \cite{boynton,newman} for
updated discussions of torsional balances). 
The torsional balance has been designed at ILL and its construction
is in progress (see Fig. 2 for a schematic layout of the apparatus).
The two pairs of plates for the Casimir force measurement are installed at
the opposite arms of the moving assembly of the balance and on the static parts
of the balance respectively.
The plates for the Casimir force measurement and for the electrostatic
calibration/actuation have a surface area of the order of $120 \mathrm{cm}^2$
for measurements at the largest gaps, around $10\mu$m. If needed, the area can 
be reduced to $15 \mathrm{cm}^2$ for measurements at smallest gaps. 
The larger mirror size increases the Casimir force but leads to a larger mass
of the plates/frame assembly, a thicker suspension wire and consequently a lower
absolute sensitivity of the force measurement.
The optimal parameters will be chosen experimentally by maximizing the
measurement accuracy while minimizing systematic effects.
Based on the measurement of the free oscillation of the balance with a
tungsten or quartz wire of diameter 50-150$\mu$m, the torque sensitivity
is estimated in the 1-100$\mu$N/rad range.
Considering the leverage of the balance and a minimum detectable displacement
of $\simeq 1$nm, we get a minimum detectable force smaller than 1pN. 
The position of each side plate is installed using three high precision
piezoelectric actuators switched in the dynamic positioning mode in
such a way that the distance between the side plate and the
corresponding balance plate is kept constant during the force
measurement, similarly to the procedure described in \cite{ref28}.
The absolute distance between them can be controlled with an accuracy of about 
$0.2\mu$m.
A much more precise read-out of variations of the distance will be performed by 
means of capacitors symmetrically located on the other side of the torsional 
balance with respect to the active surfaces.
Their design, for a gap of about $100 \mu$m, has been carefully chosen
to make their influence on the response of the torsional 
balance  negligible and ensure at the same time enough force sensitivity.
The readout will be based on the voltage signal required to keep the distance
constant, and will be measured capacitively, using a feedback servo circuit.
The balance will be placed in a vacuum chamber mounted on an antivibration table
with an active leveling system. Noise induced by the tilt should be
kept under control by optimizing the wire length, and can be further 
minimized by locating the center of force measurement at the same 
height as the suspension point of the wire, and by properly 
shaping the plates, as discussed in more detail in \cite{buttler}.
Small tilts in the relative position will at the 
leading order not affect the resulting Casimir force signal.  
\begin{figure}[b] 
\begin{center} \includegraphics[width=0.60\columnwidth,clip]{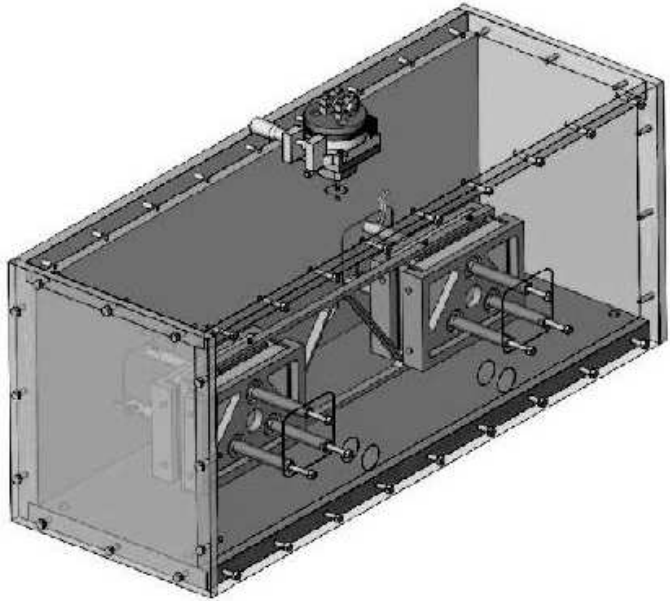} 
\caption{Torsional balance under construction at ILL, Grenoble. 
The positioning of the flat surfaces is assured through 3
piezoelectric actuators on each plate. A system of capacitors
on the other side of the balance, located at larger gaps,
allow to measure the force necessary to keep the distance
of the balance plates from the flat surfaces constant. 
The use of high-precision piezoelectric actuators and the
closed-loop option available through their controllers allow 
to push the accuracy in the parallelism down to a few times 
$10^{-7}$ radians. A four mirrors symmetric scheme allows for numerous
cross-checks on parallelism, absolute values of the gaps,
and control over many systematic effects. 
The absolute value of the gap can also be determined by using wire spacers
with diameters known within 0.2 $\mu$m.} 
\end{center} \label{fig4} 
\end{figure}

The key feature of the project is the use of parallelization
procedures in part already tested for an experiment going on
at ILL \cite{ref51,ref52,ref53,ref54,ref100}.
The experiment is aimed at observing the discretized
quantum states of ultracold neutrons in a combination of
the Earth's gravitational field and a reflecting barrier, of 10 cm size.
These states are identified through the measurement of the
neutron transmission between a horizontal mirror and an absorber.
The surfaces are at a distance of few micrometers in order
to scan the interesting region where no neutrons are
expected if they have quantized energy levels.
The quantized states have a spectrum starting with the lowest
eigenvalue at 1.4peV, which has the vertical position
probability density located mainly lower than about $15 \mu$m.
A negligible counting rate is therefore expected for smaller
separation between the mirror and the absorber.
In order to rule out systematic shifts due to distance offset, stringent 
requirements on the parallelism and the control of the separation are necessary. 

In a preliminary report \cite{ref52}, the ground state was clearly identified, later
\cite{ref100} parameters of the first and second quantum state were measured as well.
The experiment is progressing with the aim of performing precision measurements
of parameters of these and higher quantum states in order to get better
constraints for extra short-range forces \cite{ref54},
and this requires more demanding parallelism control systems.
The parallelism for sizes of the order of 10 cm is obtained by
using so-called {\it inclinometers}, which have a sensitivity to
deviations from parallelism better than $10^{-6}$ radians.
Piezoelectric actuators in a closed loop with precision inclinometers allow to reach 
and maintain horizontal surfaces at the same level of precision. 
This figure is a routine result of the most recent runs in
Grenoble on the experiment involving quantum states of neutron.
In comparison, the estimate of the parallelism for the Padova
Casimir force experiment was about $3 \times 10^{-5}$ radians. 

This extremely good control of deviations from parallelism makes it also
necessary to perform a precise analysis of the deviation from the 
flatness of the plates \cite{ref58}. 
We will use standard optically polished flat surfaces already utilized 
as the reflecting surface in the neutron experiment.
These surfaces are routinely studied with small angle scattering of x-rays,
and the root mean square roughness can be as low as $\simeq 1-2$nm for the not too 
thick plates to be used for the test of the thermal Casimir force.
For the search for hypothetical Yukawa forces, it will be
necessary to use metal layers with a thickness larger than 1$\mu$m
(see the discussion in the preceding section). 
This may lead to a degradation of the quality of their surface state, 
a problem which will be addressed by using already available techniques. 
First, the surface state can be studied for each sample by means of
x-ray scattering, allowing one to obtain the Fourier spectrum
of the deviations from perfect flatness \cite{ref58}.
Then, these data can be transformed into force corrections by using 
modelizations of nonspecular scattering recently developed for the purpose
of evaluating the effect of roughness \cite{ref59,ref59b}. 
In any event, this issue will be assessed more accurately than in previous
experiments due to the availability of surface analysis 
facilities to provide in situ characterization of the apparatus. 
The remaining concern may be the preservation of the flatness state of the 
mirrors along the time or after the mounting of the experiment.

\section{Conclusions}

In this paper, we have presented a proposal to study Casimir forces in the
distance range from a few $\mu$m to 10$\mu$m.
This distance range is the one where the theoretical predictions look the most promising. 
At the same time, it is a ``no experiment's land'' where scarce progress have been
reported so far, in contrast to the more widely explored regions of either shorter distances,
with Casimir experiments up to $\simeq 1\mu$m, or larger distances, with Cavendish type
experiments down to $\simeq 100\mu$m.

The program will consist in assembling the apparatus, checking the plate parallelism, calibrating the instrument with known signals like electrostatic forces
and studying the background and systematic effects, in particular the
residual electrostatic effects. A complete characterization of 
the apparatus' sensitivity and noise will be performed. The apparatus may be also be 
calibrated with
a physical signal such as the gravitational force, by adding controllable 
external masses as sources like in a Cavendish experiment. 
This physical calibration will bypass all the current controversies on the actual 
sensitivity of some of the current experiments on Casimir forces.  
The measurement will first aim at testing the
conflicting modelizations of thermal corrections to the Casimir force. 
In particular, these results should contribute to settle the controversy
going on in the literature by bringing into the debate experimental information. 
On a longer term, the goal will be to improve the constraints on non-Newtonian
gravity by analyzing the residuals between experimental measurements and theoretical predictions.

\begin{acknowledgments}
We thank S. Baessler, C. Binns, J. Chevrier, H. Gies, W. Lippert, 
P. Maia Neto, A.K. Petoukhov, M. Scherer,  and C. Speake for stimulating
discussions. We acknowledge the use of unpublished material 
kindly trasmitted by the authors of Ref.\cite{Coy03}.
\end{acknowledgments}


\begin{references}
\bibitem{ref8} Fischbach E and Talmadge C L 1999 {\it The search for Non-Newtonian gravity}
(AIP/Springer-Verlag, New York) and references therein.
\bibitem{ref9} Long J C, Chan H W and Price J C 1999 {\it Nucl. Phys.} \textbf{B 539} 23
\bibitem{Adelberger03} Adelberger E G, Heckel B R and Nelson A E 2003
{\it Ann. Rev. Nucl. Part. Sci.} \textbf{53} 77 [hep-ph/0307284]
\bibitem{Coy03} Coy J, Fischbach E, Hellings R, Talmadge C and Standish E M 2003,
private communication 
\bibitem{Jaekel05ijmp}  Jaekel M-T and Reynaud S 2005
{\it Int. J. Mod. Phys.} \textbf{A 20} 2294 [hep-ph/0501038]
\bibitem{ref12} Smith G L {\it et al.} 2000 {\it Phys. Rev.} \textbf{D 61} 022001
\bibitem{ref13} Hoyle C D {\it et al.} 2001 {\it Phys. Rev. Lett.} \textbf{86} 1418
\bibitem{ref13b} Hoyle C D {\it et al.} 2004 {\it Phys. Rev.} \textbf{D 70} 042004
\bibitem{ref14} Long J C {\it et al.} 2003 {\it Nature} \textbf{421} 922
\bibitem{ref15} Chiaverini J {\it et al.} 2003 {\it Phys. Rev. Lett.} \textbf{90} 151101
\bibitem{ref19} Casimir H B G 1948 {\it Proc. K. Ned. Akad. Wet.} \textbf{B 51} 793
\bibitem{ref28} Lamoreaux S K 1997 {\it Phys. Rev. Lett.} \textbf{78} 5
\bibitem{ref73} Mohideen U and Roy A 1998 {\it Phys. Rev. Lett.} \textbf{81} 4549
\bibitem{73b} Harris B W, Chen F and Mohideen U 2002 {\it Phys. Rev.} \textbf{A 62} 052109
\bibitem{73c} Chen F, Klimchitskaya G L, Mohideen U and Mostepanenko V M 2004
{\it Phys. Rev.} \textbf{A 69} 022117
\bibitem{ref74} Ederth T 2000 {\it Phys. Rev.} \textbf{A 62} 062104
\bibitem{ref49} Chan H B, Aksyuk V A, Kleiman R N, Bishop D J and Capasso F 2001 
{\it Science} \textbf{291} 1941
\bibitem{ref50} Chan H B, Aksyuk V A, Kleiman R N, Bishop D J and Capasso F 2001 
{\it Phys. Rev. Lett.} \textbf{87} 211801
\bibitem{ref56} Bressi G, Carugno G, Galvani A, Onofrio R, Ruoso G and Veronese F 2001 
{\it Class. Quant. Grav.} \textbf{18} 3943
\bibitem{ref36} Bressi G, Carugno G, Onofrio R and Ruoso G 2002 
{\it Phys. Rev. Lett.} \textbf{88} 041804
\bibitem{ref75} Decca R S, Lopez D, Fischbach E and Krause D E 2003
{\it Phys. Rev. Lett.} \textbf{91} 050402
\bibitem{ref75b} Decca R S, Fischbach E, Klimchitskaya G L, Krause D E,
Lopez D and Mostepanenko V M 2003 {\it Phys. Rev.} \textbf{D 68} 116003
\bibitem{ref16} Antoniadis I, Dimopoulos S and Dvali G 1998 {\it Nucl. Phys.} \textbf{B 516} 70
\bibitem{ref17} Arkani-Hamed N, Dimopoulos S and Dvali G 1999
{\it Phys. Rev.} \textbf{D 59} 086004
\bibitem{ref18} Arkani-Hamed N, Dimopoulos S and Dvali G 2003 {\it C. R. Physique} \textbf{4} 347
\bibitem{ref20} Plunien G, M\"uller B and Greiner W 1986 {\it Phys. Rep.} \textbf{134} 87
\bibitem{ref21} Mostepanenko V M and Trunov N N 1997
{\it The Casimir effect and its applications}, (Clarendon, London)
\bibitem{ref22} Bordag M 1999 {\it The Casimir effect 50 years later} (World Scientific, Singapore)
\bibitem{ref23} Bordag M, Mohideen U and Mostepanenko V M 2001 {\it Phys. Rep.} \textbf{353} 1
\bibitem{ref24} Reynaud S, Lambrecht A, Genet C and Jaekel M T 2001
in {\sl Space Missions for Fundamental Physics}, {\it C. R. Acad. Sci. Paris} {\bf 2-IV} 1287
\bibitem{ref24b} Lambrecht A and Reynaud S 2002 {\sl Vacuum Energy}, 
{\it S\'eminaire Poincar\'e} \textbf{1} 107
\bibitem{ref25} Milton K A 2001 {\it The Casimir Effect: 
Physical Manifestations of the Zero-Point Energy}
(World Scientific, Singapore)
\bibitem{ref26} Milonni P 1994 {\it The quantum vacuum} (Academic Press, San Diego)
\bibitem{ref24c} Genet C, Lambrecht A, and Reynaud S 2002
{\sl On the nature of dark energy} 121 (Frontier Group)
\bibitem{ref27} Milton K A 2004 {\it J. Phys.} \textbf{A 37} R209
\bibitem{Lifshitz56} Lifshitz E M 1956 {\it Sov. Phys. JETP} \textbf{2} 73
\bibitem{Mehra67} Mehra J 1967 {\it Physica} \textbf{57} 147
\bibitem{Brown69} Brown L S and Maclay G J 1969 {\it Phys. Rev.} \textbf{184} 1272
\bibitem{Schwinger78} Schwinger J, de Raad L L and Milton K A 1978 {\it Ann. Phys.} \textbf{115} 1
\bibitem{ref29} Lamoreaux S K 1998 {\it Phys. Rev. Lett.} \textbf{81} 5475
\bibitem{ref30} Lambrecht A and Reynaud S 2000 {\it Phys. Rev. Lett.} \textbf{84} 5672; 
see also {\it Euro. Phys. J.} \textbf{D 8} 309
\bibitem{ref31} Lamoreaux S K 2000 {\it Phys. Rev. Lett.} \textbf{84} 5673
\bibitem{ref31b} Lamoreaux S K 2000 {\it Comments Mod. Phys.} \textbf{1 D} 247
\bibitem{ref32} Bostrom M and Sernelius B E 2000 {\it Phys. Rev. Lett.} {\bf 84} 4757
\bibitem{ref33} Lamoreaux S K 2001 {\it Phys. Rev. Lett.} \textbf{87} 139101
\bibitem{ref34} Sernelius B E 2001 {\it Phys. Rev. Lett.} \textbf{87} 139102
\bibitem{ref60} Svetovoy V B and Lokhanin M V 2000 {\it Mod. Phys. Lett. A} \textbf{15} 1437
\bibitem{ref67} Svetovoy V B and Lokhanin M V 2003 {\it Phys. Rev. A} \textbf{67} 022113
\bibitem{ref63} Geyer B, Klimchitskaya G L and Mostepanenko V M 2002
{\it Phys. Rev.} \textbf{A 65} 062109
\bibitem{ref68} Geyer B, Klimchitskaya G L and Mostepanenko V M 2003
{\it Phys. Rev.} \textbf{A 67} 062102
\bibitem{ref66} H{\o}ye J S, Brevik I, Aarseth J B and Milton K A 2003
{\it Phys. Rev.} \textbf{E 67} 056116
\bibitem{ref35} Torgerson J R and Lamoreaux S K 2004
{\it Phys. Rev.} \textbf{E 70} 047102
\bibitem{ref37} Onofrio R 2004 {\it Proc. 6th Workshop on Quantum Field Theory under the
Influence of External Conditions}, edited by K. A. Milton (Rinton Press, Princeton, NJ)
\bibitem{ref38} Derjaguin B V 1960 {\it Sci. Am.} \textbf{203} 47
\bibitem{ref39} Blocki J, Randrup J, Swiatecki W J and Tsang C F 1977
{\it Ann. Phys.} \textbf{105} 427
\bibitem{ref51} Nesvizhevsky V V {\it et al.} 2000 {\it Nucl. Instr. Meth.} \textbf{A 440} 754
\bibitem{ref52} Nesvizhevsky V V {\it et al.} 2002 {\it Nature} \textbf{415}, 297
\bibitem{ref53} Nesvizhevsky V V {\it et al.} 2003 {\it Phys. Rev.} \textbf{D 67} 102002
\bibitem{ref54} Nesvizhevsky V V and Protasov K V 2004  {\it Class. Quantum Grav.} \textbf{21} 4557
\bibitem{ref100} Nesvizhevsky V V {\it et al.} 2005 {\it Eur. Phys. J.} \textbf{C 40} 479
\bibitem{ref61} Genet C, Lambrecht A and Reynaud S 2000 {\it Phys. Rev.} \textbf{A 62} 012110
\bibitem{ref62} Genet C, Lambrecht A and Reynaud S 2002 {\it Int. J. Mod. Phys.} \textbf{A 17} 761
\bibitem{ref70} Dalvit D A R , Lombardo F C, Mazzitelli F D and Onofrio R 2004 
{\it Europhys. Lett.} \textbf{67} 517
\bibitem{ref55} Speake C C and Trenkel C 2003 {\it Phys. Rev. Lett.} \textbf{90} 160403
\bibitem{refholger} Gies H 2005 {\sl Quantum energies and forces with
  worldline numerics}, talk at the {\it $7th$ Workshop on Quantum Field Theory under the
Influence of External Conditions, QFEXT07, Barcelona}
\bibitem{refholger1} Gies H 2005, private communication 
\bibitem{Long00} Long J C, Churnside A B and Price J C 2000 
{\it Proceedings of the Ninth Marcel Grossmann Meeting on General Relativity} 
(World Scientific) 1825 [arXiv:hep-ph/0009062]
\bibitem{boynton} Boynton P E 2000 {\it Class. Quantum Grav.} \textbf{17} 2319
\bibitem{newman} Newman R 2001 {\it Class. Quantum Grav.} \textbf{18} 2407
\bibitem{buttler} Lamoreaux S K and Buttler W T 2005 {\it Phys. Rev. E} \textbf{71} 036109
\bibitem{ref58} Genet C, Lambrecht A, Neto P M and Reynaud S 2003 {\it Europhys. Lett.} \textbf{62} 484
\bibitem{ref59} Maia Neto P M, Lambrecht A and Reynaud S 2005 {\it Europhys. Lett.} \textbf{69} 924
\bibitem{ref59b} Maia Neto P M, Lambrecht A and Reynaud S 2005
{\it Phys. Rev.} \textbf{A 72} 012115
\end{references}
\end{document}